# SoS-RPL: Securing Internet of Things Against Sinkhole Attack Using RPL Protocol-based Node Rating and Ranking Mechanism

Mina Zaminkar[1] . Reza Fotohi[2]

**Abstract**

Through the Internet of Things (IoT) the internet scope is established by the integration of physical things to classify themselves into mutual things. A physical thing can be created by this inventive perception to signify itself in the digital world. Regarding the physical things that are related to the internet, it is worth noting that considering numerous theories and upcoming predictions, they mostly require protected structures, moreover, they are at risk of several attacks. IoTs are endowed with particular routing disobedience called sinkhole attack owing to their distributed features. In these attacks, a malicious node broadcasts illusive information regarding the routings to impose itself as a route towards specific nodes for the neighboring nodes and thus, attract data traffic. RPL (IP-V6 routing protocol for efficient and low-energy networks) is a standard routing protocol which is mainly employed in sensor networks and IoT. This protocol is called SoS-RPL consisting of two key sections of the sinkhole detection. In the first section rating and ranking the nodes in the RPL is carried out based on distance measurements. The second section is in charge of discovering the misbehavior sources within the IoT network through, the Average Packet Transmission RREQ (APT-RREQ). Here, the technique is assessed through wide simulations performed within the NS-3 environment. Based on the results of the simulation, it is indicated that the IoT network behavior metrics are enhanced based on the detection rate, false-negative rate, false-positive rate, packet delivery rate, maximum throughput, and packet loss rate.

**Keywords** Internet of Things (IoT) . Sinkhole attack . RPL . Rating and ranking mechanism

✉ Mina Zaminkar
   Zaminkar@ashrafi.ac.ir

✉ Reza Fotohi*
   R_fotohi@sbu.ac.ir; Fotohi.reza@gmail.com

[1]   Department of Computer Engineering, Faculty of Engineering, Shahid Ashrafi Esfahani University, Isfahan, Iran
[2]   Faculty of Computer Science and Engineering, Shahid Beheshti University, G. C. Evin, Tehran, Iran

# 1 Introduction

There is a rising effort for connecting large physical objects in short distances on the internet utilizing the IPV6 protocols to create the internet of things. Standardizing the routing protocols of lossy and low power networks (RPL) has recently made an IoT routing protocol. RPL was originally intended to be used in lossy and low power (LLN) networks. Within RPL, a Destination Oriented Directed Acyclic Graph (DODAG) is created among the nodes in 6LoWPAN which supports one-way traffic to the destination, two-way traffic between devices, and two-way traffic between the devices and the destination. Also, IPV6 is called on over Low-Power Wireless Personal Area Networks 6LoWPAN, which is a wireless sensor network using the compressed IPV6 protocol for networking and IEEE 802.15.4 as the physical layer and data link protocol. Dissimilar to normal independent WSN networks, the devices confined in IoT are available everywhere. Therefore, they encounter attacks from both the internet and inside the network [1-2].

A physical object is able to potentially connect to the IoT utilizing IPV6. For IoT, multiple applications exist. The scope of the applications includes home security management and home automation, environmental monitoring, smart energy management, and monitoring, industrial automation, transportation tracking, security and military, smart cities, and medical supervision. In the real world, implementation of IoT requires secure connections as a significant challenge due to the heterogeneousness of IoT tools: some resources are limited while others can connect to powerful IP hosts. Also, the connections between devices in the IoT essentially need to be secure end to end (E2E) connections. This means to emphasize the integrity and confidentiality of the messages from the source to the destination. Because there are several attacks, like the sinkhole attack, providing IoT security is very important.

In this paper, the main objective is to design a defense mechanism against sinkhole attacks between IoT devices which operates using node rating and ranking. In this scenario, we tried to realistically provide IoT tools arranged in an enterprise scenery affected by real-world sinkhole, and executing genuine attacks. For replicating a characteristic structural data flow, we gathered the traffic data from IoT tools connected through Wi-Fi to various access points, the wire connected to a central switch and connected to a router as well. To sniff the network traffic, we carried out port mirroring on the switch to record the data utilizing Wireshark. We evaluated our discovery technique as realistically as possible and deployed all elements of two botnets (Fig. 1) in our isolated lab to use them for infecting 9 commercial IoT tools.

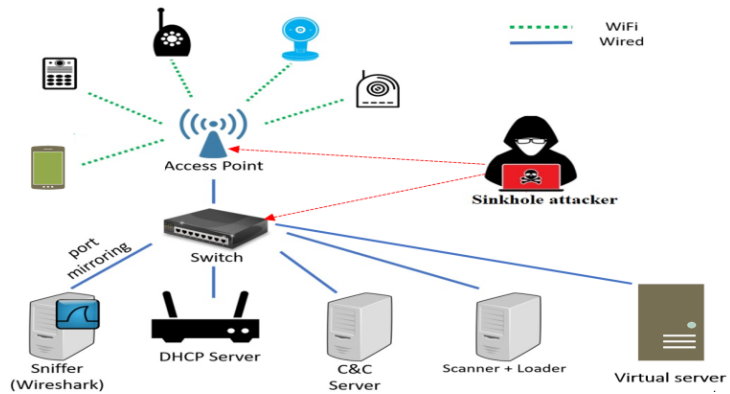

**Fig. 1** Lab setup for detecting IoT sinkhole attacks.

The rest of this study is adjusted as: Section 2 explains related work. Section 3 explains security attacks of IoT network. In Section 4, we explain the proposed method the SoS-RPL schema. Section 5 includes evaluations of simulation experiment results. Ultimately, we provide the conclusion in Section 6.

## 2 Related works

Different security measurements were established and utilized in various studies for stating the sinkhole attack and protecting IoTs from this attack. It is not a new subject, and widespread studies exist in this regard. Different studies have proposed various techniques to state these attacks.

To protect IoT sensors versus a huge amount of cyber-attacks, a methodology has been developed by Pacheco et al. [3]. At first, they presented the IoT security structure to SIs that involves 4 layers including devices (end nodes), network, services, and application. At that point, their methodology was exhibited in order to develop a general threat model for distinguishing the weaknesses in each layer and the potential countermeasures which can be spread out to diminish their taking advantage. It is worthy to note that authors indicated how it is possible an anomaly behavior analysis invasion detection system (ABA-IDS) established upon the discrete wavelet transform (DWT) develops to discover abnormalities that could be activated by means of attacks in contrast to the sensors in the first layer of our IoT structure.

A spectrum of challenges, attitudes, and practice in IoT security has been taken into account in [4]. IoT security is exclusive in several ways. Moreover, it lets us know about numerous experiments which are dissimilar from those in security guarantee of other computing devices like desktops, laptops, servers, or even mobile devices. They specially develop two classifications of security attacks with respect to the IoT system. The first one presents attacks on the four-layer structural design of IoT including perception, network, middleware, and application layer. On the foundation of that, they analytically investigated the security dangers and privacy concerns on every single layer of IoT. Because of occurring the attacks in each layer of IoT, the authors had to provide a security barrier for the whole IoT structure, not only for a particular technology. The second one of the IoT security and susceptibilities is dependent upon various application scenarios. The second classification creates a systematic basis to protect different IoT applications.

As a new public significant cryptography, Lattice-Based cryptography has been developed in [5] to substitute the public one. In order to execute lattice-based cryptography, the Ring-LWE scheme has been recommended. There should be an optimization for applying the scheme to the IoT devices via 8-bit, 32-bit, or 64-bit microcontrollers. It should be noted that the 8-bit environment is very significant for small IoT devices. Nevertheless, side-channel attacks can be damaged the Ring-LWE performance. In this research, using 8-bit microcontrollers, we analyze the attack scenarios and offer a countermeasure by bit examination for the IoT applications.

To validate and interconnect the main generation between the IoT devices, a lightweight physical-layer established security plan is recommended in [6]. They have scientifically examined and evaluated the developed method by considering the practicability of real executions. Additionally, they have comprehensively suggested a physical-layer main generation and identification scheme established upon frequency hopping communications as the RSSs of distinctive frequencies create its parameter sets.

The history, background, and statistics of IoT, also, security-based analysis of IoT architecture have been thoroughly discussed in [7]. Besides, they have provided two types of classifications including security challenges in the IoT environment and different protection mechanisms. They have also concluded that investigation on numerous research challenges, which exist in the literature yet, can provide a superior realization about the problem, present elucidation space, and upcoming research guidelines to protect the IoT versus the different attacks.

On the basis of elliptic curve cryptography (ECC), Alamr et al. [8] have recommended a new radio-frequency identification verification procedure in order to get rid of lots of weaknesses. As well, they have utilized elliptic curve Diffie–Hellman (ECDH) vital agreement procedure to generate a provisional shared key which is served to encrypt the future conveyed messages. Their procedure attains a set of security properties such as reciprocal confirmation, unrecognizability, secrecy, forward security, location privacy, and the withstanding against man-in-the-middle, replay and impersonation attacks.

The aforementioned IoT problems in the network security have been introduced and the requirement of invasion detection indicated in [9]. A number of categories of invasion uncovering technologies have been talked about and their application on IoT architecture has been studied. They have compared the application of various technologies and made a viewpoint of the following phase of research. The study of network invasion technology can be a crucial topic through data mining and machine learning approaches. More than one class feature or detection model is required to increase the exposure rate of network invasion uncovering.

The IoT security problem has been addressed in [10]. They want to obstruct the attacks at the network level rather than device one using SDN. Their goal was to defend the IoT devices from malevolent attacks and diminished the created damage. The attack is almost certainly begun by the IoT device itself or the device is the target. A framework and soft things for the IoT security established upon the SDN methods, which assists in quick recognition of unusual behavior and heightened flexibility, have been presented in [10]. They have executed the concept proof on Mininet emulator in order to distinguish irregular traffic of IoT with a Support Vector Machine (SVM), machine learning algorithm, and succeeding alleviation of the attacks. Moreover, they have taken into account lots of attacks such as TCP and ICMP flooding, DDoS, and scenarios alongside the IoT device as both target and source of attacks. We compare the linear and nonlinear SVM performance in the aforementioned scenarios for the detection of these attacks.

Rostampour et al. [11] have developed an original grouping proof procedure which can be scaled. Since the scalability is a challenge in grouping proof procedure, the reader individualistically publicizes its messages and tags in order to resolve the scalability problem in the recommended procedure. To evaluate the performance of the novel technique, they have served a 64-bit lightweight Pseudo-Random Number Generator (64-PRNG) function which satisfies the requirements of low-power and low-cost systems.

To confirm the security technology, a test bed has been fabricated to discover the potential cyber-attacks in the next-generation intelligent power control system environment which is defined like IEC and NIST in standard documents and directed the investigates to approve the appropriateness of the test bed [12]. The suggested test bed can steadily integrate the new security technologies into the industrial important substructure. Besides, it is also predictable that system security and steadiness will be improved.

The nodes mobility problem has been scrutinized so that the recommended solution has an appropriate performance in portable environments [13]. Their security mechanism is founded on the reliance concept. Reliance is a level of security that every single thing has from the other things for achievement in the demanded job without leading to security complications. To reliance things in the IoT having a multi-dimensional visualization of the reliance, they have provided a widespread hierarchical model. The three most key dimensions that should be taken into account are as follows: quality of p2p communication and service and background information. These dimensions and lively and versatile techniques, which are utilized in the calculation of the reliance and provided a mechanism in order to serve the computed reliance, make available security necessities to handle the attacks in the IoT movable environment despite the fact that network performance increases. It is worthy to mention that these dimensions are not restricted and the model has the aptitude to take into account the other ones on the foundation of the calculation purpose of the reliance. In the developed technique, they have incorporated the reliance model into RPL and provided an innovative OF. The recommended new RPL procedure was experimentally assessed under attacks of BLACKHOLE, SYBIL, and RANK in connection with subsequent performance metrics as packet loss rate, end-to-end delay, and average parent variations.

This research work is intended to implement a new methodology, i.e. profound learning, related to the cybersecurity to facilitate the attacks revealing in the public internet of things. The profound model performance has been compared to the traditional machine learning method, also, the distributed attack detection (DAD) has been assessed versus the concentrate uncovering system [14].

In the presence of three individual packets dropping attacks, a sensitivity analysis of TRS-PD preformed through a change of different parameters values in various network scenarios have been accomplished in [15]. Moreover, this work was a summary of the attack-pattern detection mechanism, reliance model, and routing mechanism adopted by TRS-PD to withstand the opponents which follow the specific attack patterns accompanied by the other ones.

Zakaria et al. [18] have impressed via the SDN abilities as they have presented a complete review of obtainable SDN-based DDoS attack uncovering and alleviation solutions. According to the DDoS attack discovery, they have categorized solutions techniques and determined the necessities of an operational solution. Furthermore, on the basis of their outcomes, they have recommended an original framework for uncovering and alleviation of DDoS attacks in a large-scale network which composes of a smart city built on the SDN substructure. Their recommended framework is able to satisfy the application-specific DDoS attack discovery and alleviation needs. The most important involvement is double. First, they have provided a detailed investigation and argument of SDN-based DDoS attack discovery and alleviation mechanisms, also, they have categorized them regarding the discovery methods. Second, by leveraging the SDN features for the network security, they have recommended and developed an SDN-established proactive DDoS Defense Framework (ProDefense).

A basis location security procedure based on dynamic routing addresses the source location confidentiality problem. The authors have introduced a self-motivated routing scheme which aims at maximizing tracks for data broadcast. At first, the suggested scheme arbitrarily selects a preliminary node from the network boundary. All of the packages will make a journey through an avaricious and successive directed route before attainment to the sink [19].

MLDMF has been presented for IIoT in [20] which comprises the cloud, fog, and edge computing level. Software-defined networking (SDN) has been utilized to manipulate the network. These two frameworks are combined to advance access security and effectual controlling of IIoT.

A method called REATO has been presented to identify and neutralize a DoS attack in contrast to the IoT middleware known as NPS. The premeditated solution tailored to the NPS architecture has been authenticated using a real test-bed and composed by a NPS sample mounted on a Raspberry Pi that receives open data feeds in real time via an adaptable set of sources. The work started from the obligation to find out a solution is capable of to guard an IoT system towards DoS attacks by considering all the potential circumstances that can take place (i.e., attacks to the data sources and attacks to the IoT platform) [21].

In [22], a deep-learning established machine learning technique is provided for the IoT to discover the routing attacks. Using the Cooja IoT emulator, attack data with high fidelity have been generated within IoT networks containing 10-1000 nodes. Here, a profound learning-based attack detecting method with high scalability was recommended for uncovering the IoT routing attacks that are reduced rank, hello-flood, and version number modifying attacks by amazing meticulousness and accurateness. To use deep learning for cyber security within the IoT, the availability of significant IoT attack data is essential.

In the article [23], a trust and time-based RPL routing protocol (SecTrust-RPL) is proposed to secure IoT networks from routing attacks. A Secure Trust (SecTrust) architecture is embedded in the RPL routing protocol to protect against malicious attacks based on trust to detect and isolate attacks in order to improve security and network performance.

In the article [24], the proposed solution consists of two stages; in the first stage, using fuzzy logic-based approach to detect on-off attacks, as well as intruders who perform destructive behaviors in the network. In the second step, the authors proposed a fuzzy logic-based method for identifying destructive nodes involved in providing destructive services.

Table 1 recapitulates the performed efforts in order to design IDS for the IoT ("-" stands for an indefinite characteristic).

**Table 1:** Comparison between detection schemes for IoT

| References | Placement schema | Detection schema | Attack type | Validation schema |
|---|---|---|---|---|
| [3] | Centralized | Anomaly-based | DoS | _Simulation |
| [4] | Hybrid | Hybrid | Routing attack | _Simulation |
| [5] | Distributed | Signature-based | Side-channel attack | _None |
| [6] | Hybrid | Hybrid | Physical-layer attack | _Simulation |
| [7] | Hybrid | Hybrid | Multiple conventional attacks | _Simulation |
| [8] | Distributed | Signature-based | MIMA, replay and impersonation attack | – |
| [9] | – | Signature-based | Multiple conventional attacks | – |
| [10] | Centralized | Anomaly-based | DDoS | – |
| [11] | – | Hybrid | RFID attacks | _None |
| [12] | Centralized | Hybrid | Cyber-attacks | _Simulation |
| [13] | | Hybrid | Routing attacks | _None |
| [14] | Distributed | Signature-based | Distributed attack | – |
| [15] | Centralized | Anomaly-based | Packet dropping attacks | – |
| [16] | – | Anomaly-based | Replay Attack | – |
| [17] | Distributed | Signature-based | Collusion attacks | _None |
| [18] | – | Anomaly-based | DDoS | – |
| [19] | Distributed | Signature-based | Cyber-attacks | _Empirical |
| [20] | – | Anomaly-based | DDoS | _Simulation |
| [21] | Centralized | Anomaly-based | DoS | _Simulation |
| [22] | Hybrid | Signature-based | Routing attacks | _Simulation |
| [23] | Distributed | Security Trust | Sybil & Rank attack | _Simulation |
| [24] | Hybrid | Fuzzy logic | On-off attacks | _Simulation |

In Table 2, a comparison of detected attacks and categories in the literature is highlighted.

**Table 2** Security threats detection schemes for IoT.

| Proposed system | Detected attacks | Category |
|---|---|---|
| Pacheco et al. (2017) | variety of cyberattacks | DoS |
| Chen et al. (2018) | Network layer attacks | Routing attack |
| Moon et al. (2018) | Side-channel and power analysis attack | Side-channel attack |
| Jiang et al. (2018) | physical-layer security | Physical-layer attack |
| Adat et al. (2017) | An energy consumption model for detecting of the DoS | Multiple conventional attacks |
| Alamr et al. (2016) | MIMA, replay and impersonation | MIMA, replay and impersonation |
| Deng et al. (2018) | hijack attack | Multiple conventional attacks |
| Bhunia and Gurusamy (2017) | malicious attacks | DDoS |
| Rostampour et al. (2017) | RFID attacks | RFID attacks |
| Lee et al. (2017) | Stuxnet attack | Cyber-attacks |
| Qin et al. (2019) | DDoS | DDoS |
| Hashemi et al. (2018) | Cyber-attack | Routing attacks |
| Diro and Chilamkurti (2017) | Topology attacks on RPL | Distributed attack |
| Jhaveri et al. (2018) | Sinkhole and neighbor attacks | Packet dropping attacks |
| Jan et al. (2017) | Cyber-attack | Replay Attack |
| Yaseen et al. (2017) | Packet forwarding misbehavior | Collusion attacks |
| Bawany et al. (2017) | DDoS | DDoS |
| Han et al. (2017) | eavesdropping, hop by and direction-oriented attack | Cyber-attacks |
| Yan et al. (2018) | multi-level DDoS | DDoS |
| Sicari et al. (2018) | Denial of Service (DoS) attack | DoS |
| Yavuz et al. (2018) | Cyber security | Routing attacks |

## 3 Security attacks of IoT network

Four groups of attacks on the IoT network are shown in Figure 2. From each group, we described attacks that were more destructive than all of those attacks.

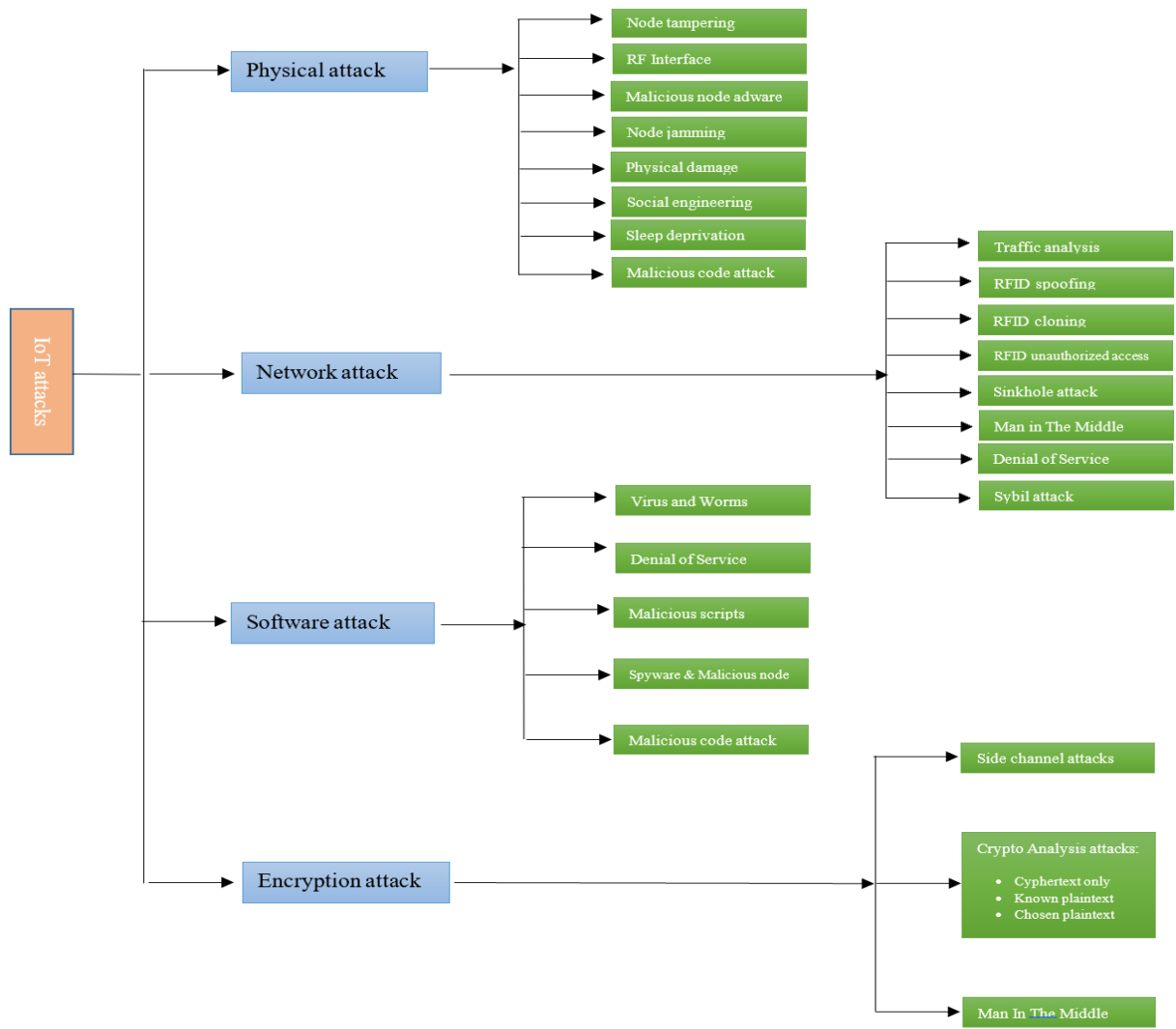

**Fig. 2** Category of attacks on IoT.

## 3.1. Physical attacks

Physical attacks carry out malicious operations on hardware devices.

1) *Node Tampering:* Physical attacks carry out malicious operations on hardware devices.

2) *RF Interference:* In this attack, the intruder performs DoS attacks by sending noisy signals on the radio frequency signals. These signals are used to communicate RFID.

3) *Node Jamming:* In this attack, the attacker disrupts the IoT wireless communication network and leads to a DoS attack.

4) *Physical Damage:* In this attack, the attacker physically damages the components of the IoT ecosystem and leads to a DoS attack.

5) *Social Engineering:* In this attack, the intruder physically contacts the users of the IoT ecosystem and manipulates their data. The intruder obtains sensitive information from the IoT to achieve its goals.

## 3.2. Network attacks

This group of attacks on the IoT ecosystem network is carrying out destructive operations.

1) *Traffic Analysis Attacks:* The hacker tracks and checks the messages exchanged on the network to obtain network information.

2) *RFID Spoofing:* The attacker deceives the RFID signals. It then records the information transmitted from an RFID tag. This type of attack seems to have the correct information and is accepted by the system.

3) *RFID Cloning:* The hacker copies data from the previous RFID tag to another RFID tag. The original ID does not copy the RFID tag. An intruder can mislead data or control data passed through a cloned node.

4) *RFID Unauthorized Access:* An intruder can view, modify, or delete information about nodes in the IoT network if the correct authentication is not performed on RFID systems.

5) *Man in the Middle Attacks:* In this attack, the enemy stopped the connection between the two nodes in the IoT network through the Internet and stole sensitive information by eavesdropping.

6) *Sinkhole attack:* One of the main attacks threatening the IoT is the attack known as the Sinkhole (SH) attack. In these attacks, a malicious node broadcasts illusive information regarding the routings to impose itself as a route towards specific nodes for the neighbouring nodes and thus, attract data traffic. The objective of this process is to draw all the traffic in the network towards the sinkhole node and as a result, alter the packets of data or silently drop them altogether. Sinkhole attacks can increase the network overhead, increase the consumption of energy and decrease the life time of the network , and ultimately annihilate the network [2]. A screenshot of a simulated RPL network is shown in Fig. 2.

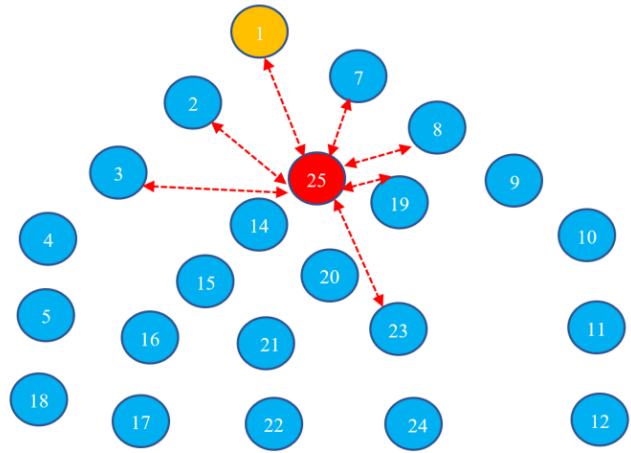

**Fig. 2** An RPL network's screenshot simulated with sinkhole attack by node 25, implementing actual IoT technologies, indicated that RPL is affected by sinkhole attacks.

## 3.3. Software attacks

In this category of attacks, the intruder uses worms, viruses, annoying advertising tools and spyware to repel services and steal data.

1) *Malicious Scripts:* In this attack, the intruder can inject the malicious script into the IoT network and take control of the system.

2) *Denial of Service:* The attacker blocks users from the Application layer and refuses to provide services to them.

## 3.4. Encryption attacks

In this type of attack, the intruder tries to get the private key to destroy the encryption technique.

1) *Side-channel Attacks:* In this attack, an intruder uses the side channel to obtain any information from the network. The attacker uses this information to identify the encryption key. This is neither a plaintext nor a ciphertext, but it does contain information about the power, the time required to perform the operation, and the error frequency.

2) *Cryptanalysis Attacks:* The attacker deceives the RFID signals. It then records the information transmitted from an RFID tag. This type of attack seems to have the correct information and is accepted by the system.

# 4 The proposed SoS-RPL schema

In the following section, we design an SoS-RPL schema by employing the rating and ranking mechanism. Three phases are contained in the SoS-RPL schema: in Sect, 3.1. The overview of the SoS-RPL schema is discussed, Sect 3.2. presents detection of the sinkhole node, and elimination of the sinkhole nodes in SoS-RPL is discussed in Sect. 3.3.

### 3.1 Phase 1: Overview of the SoS-RPL model for detecting sinkhole node

In sinkhole attacks, the malicious node declares an artificial useful route. This way, many of the nodes close to the route traffic are attracted. This attack does not disrupt network operation on its own. However, these attacks become very dangerous when they are coupled with another attack. In this section, we present an efficient security approach based on the RPL protocol. The proposed method is presented under the SoS-RPL and is designed in two phases. In the following section, we describe the proposed method.

### 3.2 Phase 2: Detection of the sinkhole node

This phase consists of two basic steps to detect sink attack.

**Step1:** The proposed SoS-RPL algorithm is implemented on top of the RPL protocol. In this algorithm, we try to use the behavior of the nodes in the network to identify and delete sinkhole nodes so that we can prevent the malicious nodes from providing false information to other nodes in the network. When the number of sinkhole nodes in the network increases, the number of transmitted DIO messages sent also increases so that the malicious node can introduce itself as the root node. Therefore, the overhead increases as the number of malicious nodes in the network increases. The higher the overhead, the higher the latency will be. Since latency is an important issue while sending security messages in IoT, overhead and therefore latency can be reduced by identifying sinkhole nodes in the network.

The following rules are used in the proposed SoS-RPL algorithm for the detection of the malicious sinkhole node:

- Any node which sends its real rank to its neighbors cannot be malicious.
- A node which has sent multiple DIO messages to the neighbors which are not its children might be malicious.
- A malicious node is a node which has misrepresented its rank.

In order to prevent routing loops, the RPL protocol calculates the hop count from a node to the root (DODAG). The variable rank represents the position of the node in the proposed method. Rank is higher if a node is farther away from the root node. Rank contains useful information for estimation of the distance from the root node. The RPL protocol provides a new control message for exchange of the routing graph information. This message, known as DIO, is used to announce the information used for the creation of DODAG. Therefore, the proposed system uses the rank value for the detection of suspicious rank values in the DIO message. The RPL protocol provides a new ICMPv6 control message for exchange of the routing graph information. The attacker can send a fake ICMPv6 routing packet in order to create a sinkhole.

In order to present the overall notion of our suggested technique, Figures 3 and 4 demonstrate how the rank values change after and prior to the sinkhole is created. Figure 3 presents how the rank value for each node is determined by RPL. The root node possesses rank zero and the rank value for each node is equal to the hop count from that node to the root plus one.

Figure 4 demonstrates how the rank value changes after the malicious node is created. M1 is deployed among the nodes and creates the sinkhole.

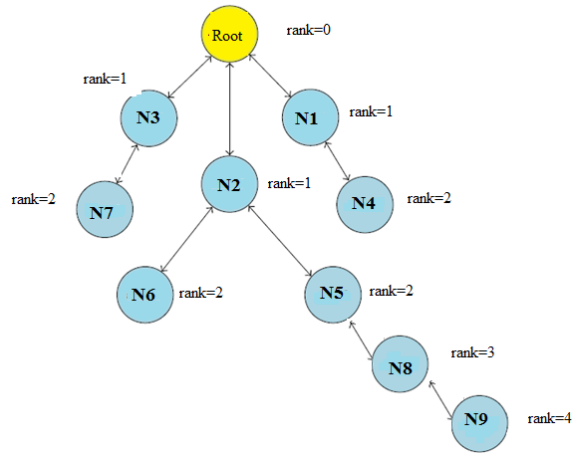

**Fig. 3** RPL protocol operation under normal conditions

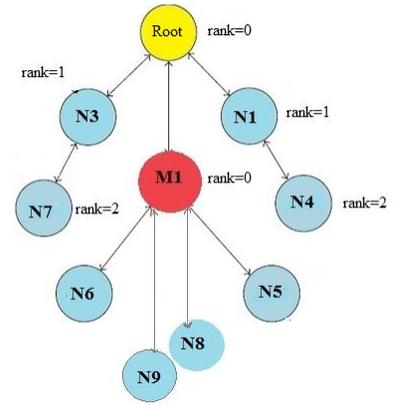

**Fig. 4** RPL protocol performance under sinkhole attack conditions

When a new node is added to the network, the root node sends a DIO message to nodes M1 and N1 in order to update the routing table. The rank values of nodes N3, N1, and M1 are one. However, when the malicious node M1 forwards the DIO message to other neighboring nodes to update the rank values, it declares its own rank to be zero. The proposed SoS-RPL method utilizes the rank value in order to identify the sinkhole node. In the proposed SoS-RPL method, DIO messages are gathered through the suggested system. The rank values are then extracted from these DIO messages. By extracting the DIO messages, the suggested system detects whether the DIO message is from a malicious node or not.

In the proposed SoS-RPL method, it is assumed that an IoT network does not contain any malicious nodes when it is deployed. The correct routing tables for every node are broadcast in the newly deployed network before the sinkhole attacks occur. The proposed method defines two features for irregular DIO message detection: DV-RANK and DI-RANK. DV-RANK is presented in Eq. (1) as the rank difference between a node and its parent. The value of this feature is attained when creating or updating the routing table:

$$DV-RANK = |(ParentRank) - (NodeRank)| \qquad (1)$$

For example, in figure 3 the DV-RANK value for node N9 is equal to one. This is because its rank is four while the rank of its parent, node N8, is equal to three. Therefore, DV-RANK of node N9 is equal to: |3-4|=1.

DI-RANK is defined as the rank difference between the node and the source node which has sent the message. This feature is calculated according to Eq. (2). For example, node N9 received the message from node N8 in figure 3. Node N9 then calculated the DI-RANK value. Since the rank of the message sender node, node N8 in this example, is three, the DI-RANK value will be equal to one. This is because the rank value of the sender node is three and the node itself has a rank equal to four. |3-4|=1.

However, when a malicious sinkhole node infiltrates the network, as represented in figure 4, this node declares its rank to be zero in the DIO message. This way, it introduces itself as the root node and sends this message to all of the neighboring nodes. When node N9 receives a DIO message from the malicious node M1, it needs to calculate the DI-RANK value. Since the rank for the source node which has sent the message is zero and the rank value for node N9 is four, DI-RANK will be equal to 4. Therefore, the DI-RANK value for node N9 in figure 4 is equal to |0-4|=4.

$$DI - RANK = |(Rank\ of\ the\ source\ message\ sender\ node) - (Rank\ of\ the\ node)| \qquad (2)$$

The proposed SoS-RPL method considers the DIO message to be malicious when DI-RANK>DV-RANK. In figure 4, the DIO message sent by node M1 will be identified as malicious by node N9 using DI-RANK>DV-RANK.

**Step2: The mechanism to detect physical layer attack in LSFA-IoT**

In the second step of our proposed SoS-RPL method, the malicious nodes producing fake $RREQ$ packets in the network are detected. As mentioned earlier, to run this step, we should first detect the misbehaviors in the network. The separation of these two steps has made the operations required for malicious node detection optimized. After running the detection process, every node should search the list of its neighbors to find the neighbor that has produced a large number of $RREQ$ packets. To detect the source of flooding attacks, each node calculates the number of produced $RREQ$ s. To do this, we use a weighted average formula in the SoS-RPL. Average Packet Transmission $RREQ$ ($APT-RREQ$) is used to calculate the average transmission of $RREQ$ packets. The average transmission is used by series data in a certain period to smooth the specified short-term and long-term fluctuations. We analyze our observations about $RREQ$ packets in a period using these calculations. $APT-RREQ$ may be calculated recursively for X series. Eq. (3) demonstrates the calculation.

$$\begin{cases} S_1 = X_1 & for\ t = 1 \\ S_t = \alpha * X_t + (1-\alpha) * S_{t-1} & for\ t > 1 \end{cases} \qquad (3)$$

Where: $\begin{cases} \alpha\ is\ a\ smoothing\ factor\ (as\ a\ constant\ value\ between\ 0\ and\ 1). \\ X_t\ is\ the\ value\ of\ RREQ\ in\ period\ t. \\ S_t\ is\ a\ value\ of\ APT-RREQ\ in\ each\ period\ t. \end{cases}$

According to the proposed method, we use different values of $\alpha$ to detect flooding attacks. The $APT-RREQ$ can be applied with the low values of $\alpha$ to check network when it is under a flooding attack. However, the high values of $\alpha$ can help to analyze the general observations of the network in a certain period and detect attack source. The number of sent $RREQ$ is determined for each node after information acquisition using the *Hello* message. Each node calculates $APT-RREQ$ value for its neighbor nodes by receiving a *Hello* message from them and getting the information of the neighbor node. We consider a threshold for $APT-RREQ$ each time. If the value of $APT-RREQ$ or a node exceeds the threshold, it indicates that the number of the $RREQ$ messages transmitted by this node is far more than the expected threshold. Therefore, this node is detected as malicious.

**3.3 Phase 3: Elimination of the sinkhole nodes**

By applying the proposed system to all of the nodes, the nodes will ignore any irrational DIO message and add the ID of the malicious node to their blacklist. Also, the nodes which have identified the malicious node, forward its ID to the root using the ICMPv6 message. This way, the root node can notify all of the nodes in the network of existence of the malicious node. Therefore, sinkhole attacks will be prevented. The proposed system is easily implemented and does not require any complex calculations or additional hardware.

Flowchart of the malicious node detection process in the proposed SoS-RPL method is presented in figure 5. Since this is a distributed algorithm, each node in the IoT network checks the properties extracted from the header packet to observe the presence of a sinkhole node within the network. As mentioned before, in order to make the detection process shorter, malicious nodes are stored in blacklists once detected. Therefore, the detection system does not need to examine them again. If the ICMPv6 message is assumed to be safe, the receivers will update their routing and neighboring nodes table.

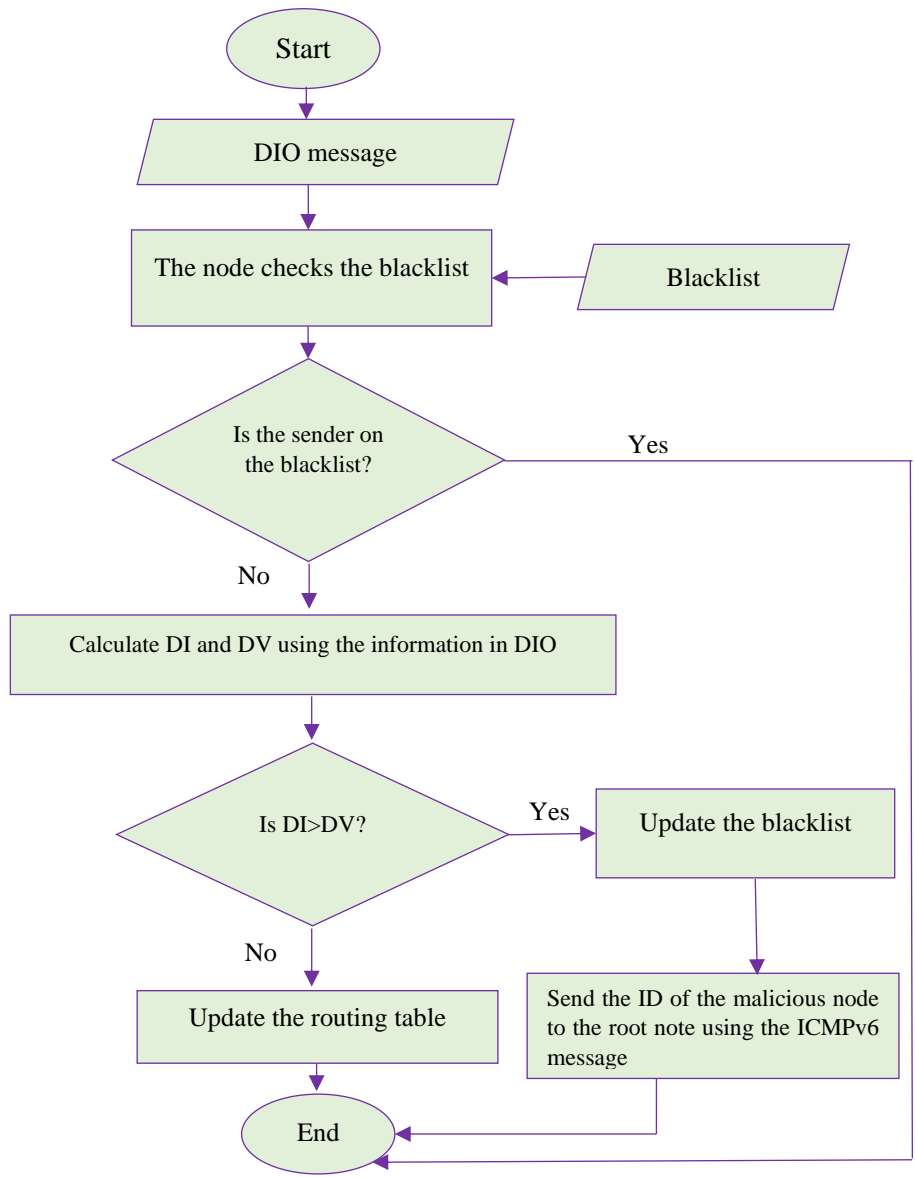

**Fig. 5** Flowchart of the SoS-RPL.

# 5 Performance evaluation

The suggested SoS-RPL performance will be assessed in the next section to avoid the sinkhole attack problem.

## 4.1 Performance metrics

The performance and efficacy of the proposed SoS-RPL method is completely investigated in this section using comprehensive simulations. The obtained results will be compared with SecTrust-RPL, Fuzzy-IoT, IRAD, and REATO methods discussed in [23], [24], [25], and [26]. The detection rate, false-negative rate, false-positive rate, packet delivery rate, maximum throughput, and packet loss rate are assessed. The meaning of notations and abbreviated used in the equations are given in Table 3 and Table 4.

**Table 3** The parameters specified for *PDR*

| Notations | Means |
|---|---|
| $X_i$ | Number of packets received by node i |
| $Y_i$ | Number of packets sent by node i |
| n | Experiments |

**Table 4** Abbreviated notations

| Parameters | Description |
|---|---|
| *FPR* | *False positive rate* |
| *FNR* | *False negative rate* |
| *TPR* | *True positive rate* |
| *TNR* | *True negative rate* |
| *DR* | *Detection rate* |
| *PDR* | *Packet delivery rate* |
| *PLR* | *Packet loss rate* |
| $A_i$ | Received packets by node i |
| $E_s$ | Packet size |
| $D_P$ | Simulation stop time |
| $D_T$ | Simulation start time |
| *Ex* | Experiments |
| $\sigma$ | Throughput |
| N | Experiments |

PDR: The number of packets received by the destination node is divided by the number of packets sent by the source node. Eq. (4) determines the PDR.

$$PDR = \frac{1}{n} * \frac{\sum_{i=1}^{n} X_i}{\sum_{i=1}^{n} Y_i} * 100\% \qquad (4)$$

FPR: The FP is determined by the total number of nodes mistakenly found as the malevolent nodes divided by the total number of usual nodes [25-27]. Hence, Eq. (5) illustrates the

$$FPR = \left(\frac{FPR}{FPR+TNR}\right)*100 \qquad \text{Where:} \qquad TNR = \left(\frac{TNR}{TNR+FPR}\right)*100 \qquad (5)$$

FNR: The rate of the malevolent node to total normal nodes incorrectly signed as a normal node [28-30]. The calculation is proved by Eq. (6).

$$FNR = \left(\frac{TPR+TNR}{All}\right)*100 \qquad \text{Where:} \qquad TPR = \left(\frac{TPR}{TPR+FNR}\right)*100 \qquad (6)$$

DR: Ratio of sinkhole nodes to total malicious nodes that were correctly diagnosed as sinkhole attack. Eq. (7) determines the DR.

$$DR = \left(\frac{TPR}{TPR+FNR}\right)*100 \qquad \text{where} \qquad All = TPR+TNR+FPR+FNR \qquad (7)$$

PLR: Percentage of data packets deleted by the intruder node [31,32]. The PLR is calculated using Eq. (8) as follows:

$$PLR = \left(\frac{1}{n}\right)*\left(\frac{\sum_{i=1}^{n}Y_i - \sum_{i=1}^{n}X_i}{\sum_{i=1}^{n}Y_i}\right)*100 \qquad (8)$$

Maximum throughput: In the IoT ecosystem, throughput is the amount of data packets generated by the source node and successfully received at the destination node. The unit of throughput is kilobits per second (Kbps) [33]. The throughput is calculated using Eq. (9) as follows:

$$\sigma = \left(\frac{1}{Ex}\right)*\left(\frac{\sum_{i=1}^{n}A_i * E_s}{D_p - D_T}\right)*\left(\frac{8}{1000}\right) \qquad (9)$$

### 4.2 The simulation environment

Since implementing and debugging IoTs in real networks is difficult, considering simulations as a basic design instrument is necessary. The primary benefit of simulation is simplification of analysis and verification of protocol, especially in large systems [34]. In this part, the suggested method's performance is assessed by NS-3 as the simulation instrument, and then the results will be discussed.

It should be noted that all SoS-RPL, REATO and IRAD settings and parameters are considered as equal.

### 4.3 Simulation results

The SoS-RPL performance is analyzed in this section under the four scenarios (Table 5). Initially, 500 nodes are deployed in the IoT area in a uniform manner. Table 5 gives some major parameters.

**Table 5:** Parameters used.

| Parameters | Value |
|---|---|
| MAC | 802.11. b |
| Traffic | CBR |
| Speed | 150 m/s |
| Size of packet | 512 Byte |
| Malicious rate | 10%, 20%, 30% |
| Type of attacks | Sinkhole |
| Transmission range | 20 M |
| Selection of target node | Random |

Table 6 gives some major parameters for four scenarios.

**Table 6** Parameters used for four scenarios.

| Scenario #1 | | Scenario #2 | |
|---|---|---|---|
| Sinkhole rate | 10% | Sinkhole rate | 20% |
| Topology (m x m) | 100*100 | Topology (m x m) | 100*100 |
| Time | 1000 | Time | 1000 |
| Scenario #3 | | Scenario #4 | |
| Sinkhole rate | 30% | Attack interval | 0, 0.05, 0.10, 0.15, 0.20, 0.25, 0.30 |
| Topology (m x m) | 100*100 | Topology (m x m) | 100*100 |
| Time | 1000 | Time | 1000 |

Performance of SoS-RPL, REATO, and IRAD in terms of DR, FNR, FPR, and PDR are shown in Table 7-10.

**Table 7** DR (30% sinkhole node of overall nodes) of various approaches.

| Attack interval | DR (%) | | | | |
|---|---|---|---|---|---|
| | IRAD | REATO | Fuzzy-IoT | SecTrust-RPL | SoS-RPL |
| 0.5 | 68 | 73 | 75 | 75 | 89 |
| 1 | 68 | 74 | 76 | 77 | 91 |
| 1.5 | 69 | 74 | 77 | 79 | 93 |
| 2 | 71 | 75 | 79 | 81 | 94 |
| 2.5 | 72 | 77 | 80 | 83 | 95 |
| 3 | 72 | 78 | 81 | 85 | 97 |
| 3.5 | 73 | 80 | 82 | 87 | 97 |
| 4 | 74 | 81 | 83 | 89 | 98 |

**Table 8** FNR (30% sinkhole node of overall nodes) of various approaches.

| Attack interval | FNR (%) | | | | |
|---|---|---|---|---|---|
| | IRAD | REATO | Fuzzy-IoT | SecTrust-RPL | SoS-RPL |
| 0.5 | 18.6 | 15.6 | 15.1 | 15 | 12.98 |
| 1 | 17.87 | 14.6 | 14.2 | 14 | 12.8 |
| 1.5 | 17.4 | 13.7 | 13.1 | 12.5 | 11.8 |
| 2 | 16.8 | 12.7 | 12.2 | 12 | 11.5 |
| 2.5 | 16.67 | 12.4 | 12 | 11 | 10.5 |
| 3 | 16.3 | 11.8 | 11.1 | 10.5 | 9.8 |
| 3.5 | 15.8 | 11.3 | 11 | 10.1 | 9.65 |
| 4 | 14.5 | 10.4 | 10.1 | 10 | 9.2 |

**Table 9** FPR (30% sinkhole node of overall nodes) of various approaches.

| Attack interval | FPR (%) | | | | |
|---|---|---|---|---|---|
| | IRAD | REATO | Fuzzy-IoT | SecTrust-RPL | SoS-RPL |
| 0.5 | 26.1 | 23.6 | 21.5 | 19.5 | 15.6 |
| 1 | 25.4 | 22.87 | 20.3 | 19.1 | 14.6 |
| 1.5 | 24.5 | 21.4 | 19.9 | 18 | 14.2 |
| 2 | 24.3 | 20.8 | 18.7 | 17 | 13.7 |
| 2.5 | 24.1 | 19.67 | 18.1 | 16 | 13.4 |
| 3 | 23.5 | 18.9 | 17.4 | 15.8 | 13.8 |
| 3.5 | 22.7 | 18.8 | 17.1 | 15.4 | 13.3 |
| 4 | 21.6 | 17.5 | 16.05 | 15 | 12.4 |

**Table 10** PDR (30% sinkhole node of overall nodes) of various approaches.

| Attack interval | PDR (%) | | | | |
|---|---|---|---|---|---|
| | IRAD | REATO | Fuzzy-IoT | SecTrust-RPL | SoS-RPL |
| 0.5 | 62 | 78 | 75 | 76 | 80 |
| 1 | 63 | 79 | 77 | 78 | 82 |
| 1.5 | 63 | 80 | 80 | 81 | 86 |
| 2 | 65 | 82 | 83 | 85 | 89 |
| 2.5 | 68 | 85 | 87 | 89 | 92 |
| 3 | 69 | 86 | 90 | 92 | 94 |
| 3.5 | 71 | 86 | 91 | 93 | 97 |
| 4 | 73 | 87 | 92 | 94 | 98 |

**DR:** Figure 6 provides a comparison between the SoS-RPL suggested scheme, REATO, IRAD, Fuzzy-IoT, and SecTrust-RPL models based on DR. Based on the diagrams, the detecting rate in every 3 approaches is reduced in terms of scenarios, particularly with the high number of attacks. For the IRAD, this reduction is much higher compared to the other mechanisms. By the suggested design, it is possible to detect all the above attacks at a detection rate of over 95%. This finding is attained by the rate of malicious nodes and the number of normal nodes as 30% and 600, respectively. The proposed design is superior as a result of the fast detecting the malicious nodes and removing them by mapping the antigenic and unsafe routes detected by the antibody-trained model and eliminated from the operation cycle. Therefore, according to Figure 6, the detection rate of the proposed method is 17%, 15, 13 and 10% higher than the REATO, IRAD, Fuzzy-IoT, and SecTrust-RPL methods, respectively.

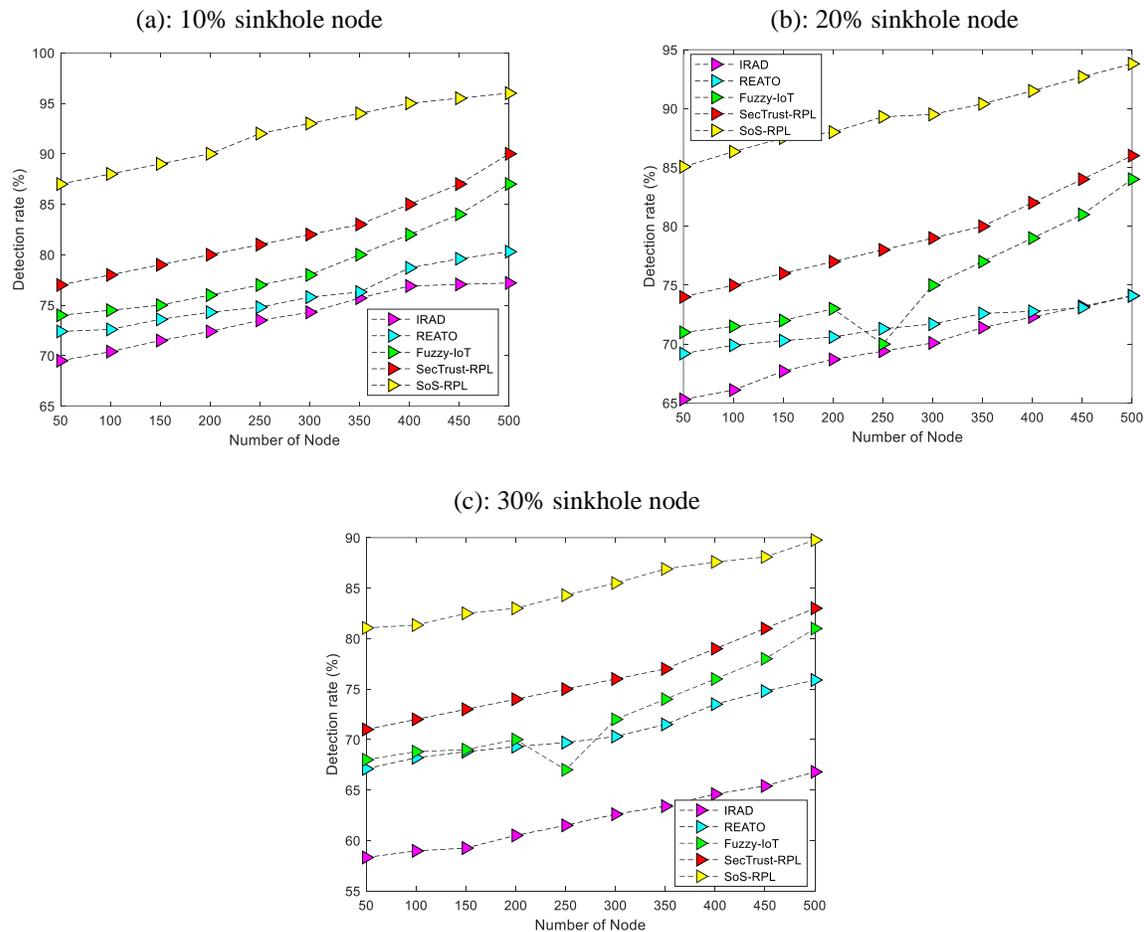

**Fig. 6** DR vs Number of nodes.

**FNR:** Figure 7 represents a comparison between the SoS-RPL suggested scheme, REATO and IRAD models based on FNR in lethal attacks. According to the diagrams, the FNR of the SoS-RPL suggested scheme incremented slightly, however, this value is much higher in the REATO and IRAD. In Figure 7(a), the suggested scheme contains the FNR of less than 1.2% by the number of normal nodes of 600, however, it is 13 and 18% respectively for the other two methods. In Figure 7(b), by the malicious nodes rate of 16%, it is less than 4% in the suggested design that is 22% and 7% for the other two approaches respectively. In Figure 7(c) the FNR is explained under security threats with the 23% malicious node). Based on the results in Figure 7(c), it is indicated that in the conventional method, over security threats, the FNR at Misbehaving nodes ratio of 0.09 is around

1.6% increasing to around 8% at 0.36 in a misbehaving nodes ratio condition. Therefore, according to Figure 6, the FNR of the proposed method is 21%, 19, 17 and 15% higher than the REATO, IRAD, Fuzzy-IoT, and SecTrust-RPL methods, respectively.

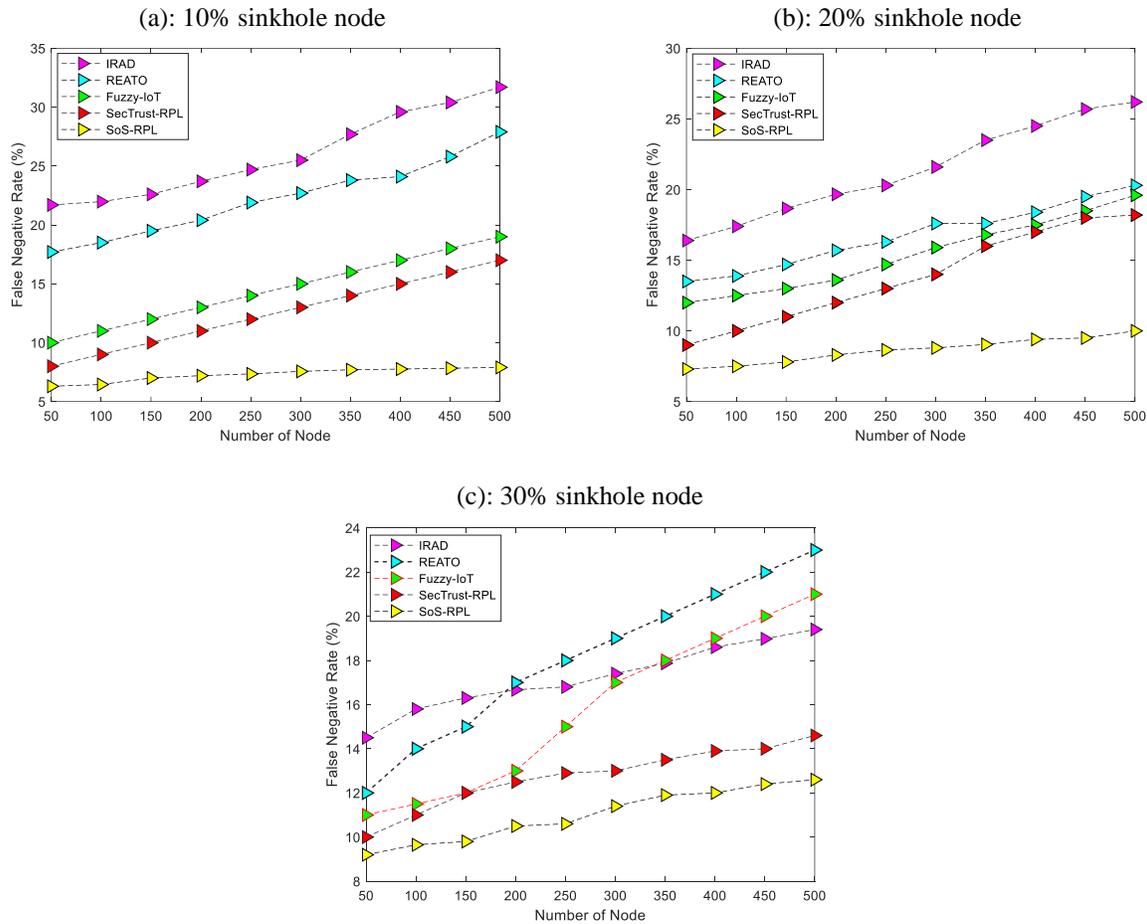

**Fig. 7** FNR vs Number of nodes.

**FPR:** in Figure 8, the comparison is provided between the suggested SoS-RPL framework against four methods of one risk-based algorithm and game theory-based techniques. According to the Figure 8(a), by the number of normal nodes within the range of 100-400 and increasing the rate of malicious nodes from 0 to 30%, a slight and moderate growth existed in the FPR created by the suggested design in comparison to the other two designs. By the malicious nodes rate and the number of normal nodes of 30% and 400, respectively, the FPR of the SoS-RPL is less than 3%. Though, this quantity is 32% for the IRAD, 28% for the REATO, 25% for the Fuzzy-IoT, and 20% for the SecTrust-RPL. The proposed design is superior as a result of its fast detection of malicious nodes and removing them by collaboration between normal nodes and ground stations that the process is carried out by the trained rules stored in memory. Moreover, it is superior by the fact that the suggested algorithm discovers the security threats and separates them from the IoT network, hence, the FPR caused by the attacks is reduced.

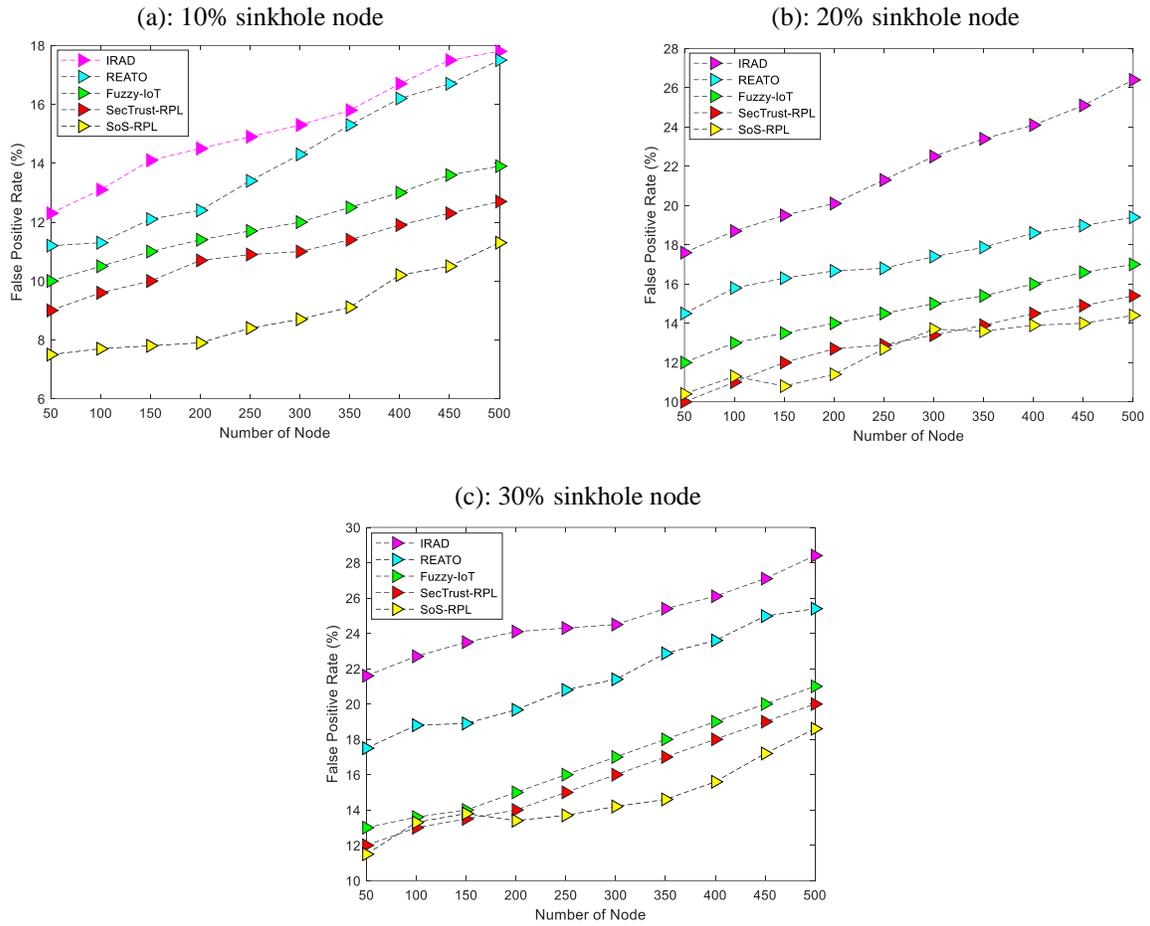

**Fig. 8** FPR vs Number of nodes.

In Figure 9, the association between the number of nodes and PDR is represented. By the number of nodes as 50, the PDR of REATO, IRAD, Fuzzy-IoT, and SecTrust-RPL are relatively low, since some packets are not able to reach the destination prior to expiring the timeout period. By incrementing the number of nodes, most packets can be delivered to the destination, hence, a slight enhancement is observed in the PDR. A slight degradation is observed in the packet delivery ratio of SoS-RPL, by the number of nodes as 50 and 100 appearing due to random factors in simulation. Based on the overall trend, SoS-RPL outperforms both REATO and IRAD based on the PDR by the number of nodes exceeding 150-500. According to Figure 9(a), (b) and (c), SoS-RPL reduces the PDR by over 32% and 22% compared to the REATO and IRAD models, respectively.

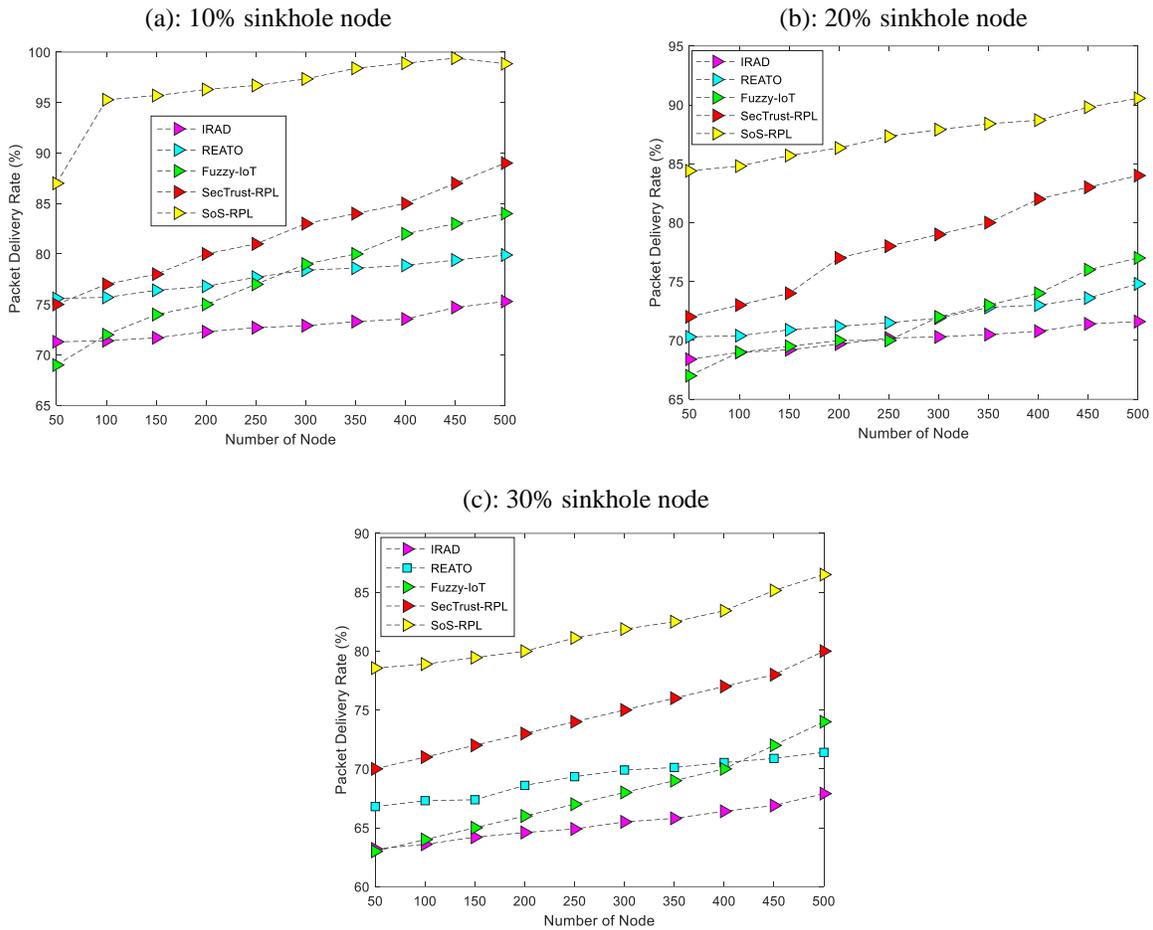

**Fig. 9** PDR vs Number of nodes.

One of the important and effective criteria in IoT ecosystem is the throughput criterion. Therefore, we considered this criterion as well for evaluating the SoS-RPL proposed method. Figure 10 shows that the proposed SoS-RPL method performs better compared to the REATO, IRAD, Fuzzy-IoT, and SecTrust-RPL protocol under the sinkhole attack at different times. This is because the proposed method makes more packets reach the destination in a single time unit. Therefore, the SoS-RPL method has higher throughput compared to the REATO, IRAD, Fuzzy-IoT, and SecTrust-RPL protocol under attack in a specified time unit. Using the ranking of the nodes along with sending security messages, malicious nodes are quickly identified in the proposed method. This leads to increased throughput and decreased delay and ultimately, increased throughput in the network.

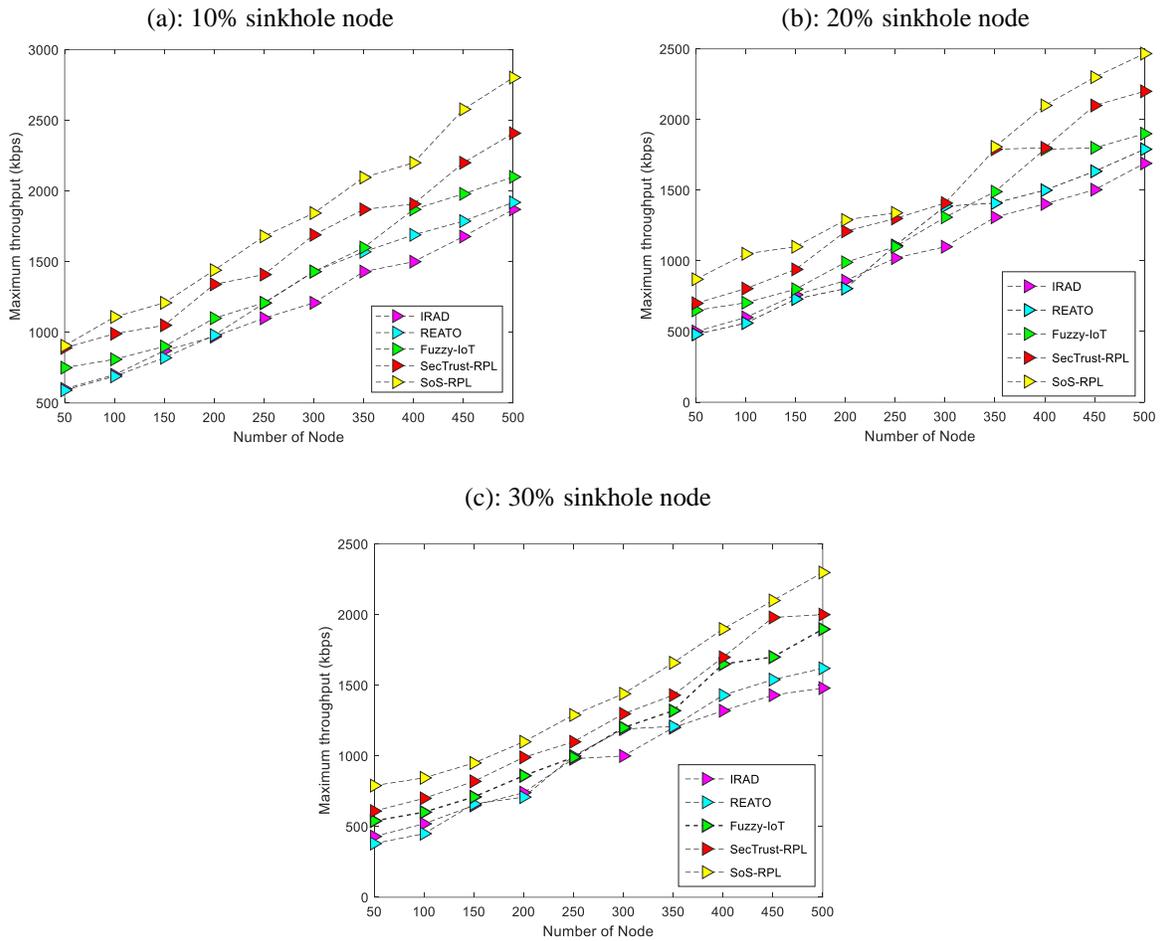

**Fig. 10** Maximum throughput vs Number of nodes.

Figure 11 illustrates the relationship between the PLR and the number of things under the identical setting expressed in Table 5 and Table 6. When the number of things is 50, we see that some packets cannot arrive themselves to the destination before the timeout period terminates; so, the packet loss ratios of REATO, IRAD, Fuzzy-IoT, and SecTrust-RPL are somewhat high. As the number of things increases, most packets can be delivered to the destination; hence, we can see a small enhancement in the packet loss ratios. The packet loss ratio of SoS-RPL has an insignificant degradation when the number of Things is 50 and 100. This is due to the presence of random factors in the simulation process. From a general point of view, when the number of things goes beyond 100-500, SoS-RPL doing better than REATO, IRAD, Fuzzy-IoT, and SecTrust-RPL methods in terms of the packet loss ratio. As shown in the Figure 11(a), (b) and (c), SoS-RPL decreases the PLR by more than 27, 15, 23, and 19% those of REATO, IRAD, Fuzzy-IoT, and SecTrust-RPL models, respectively.

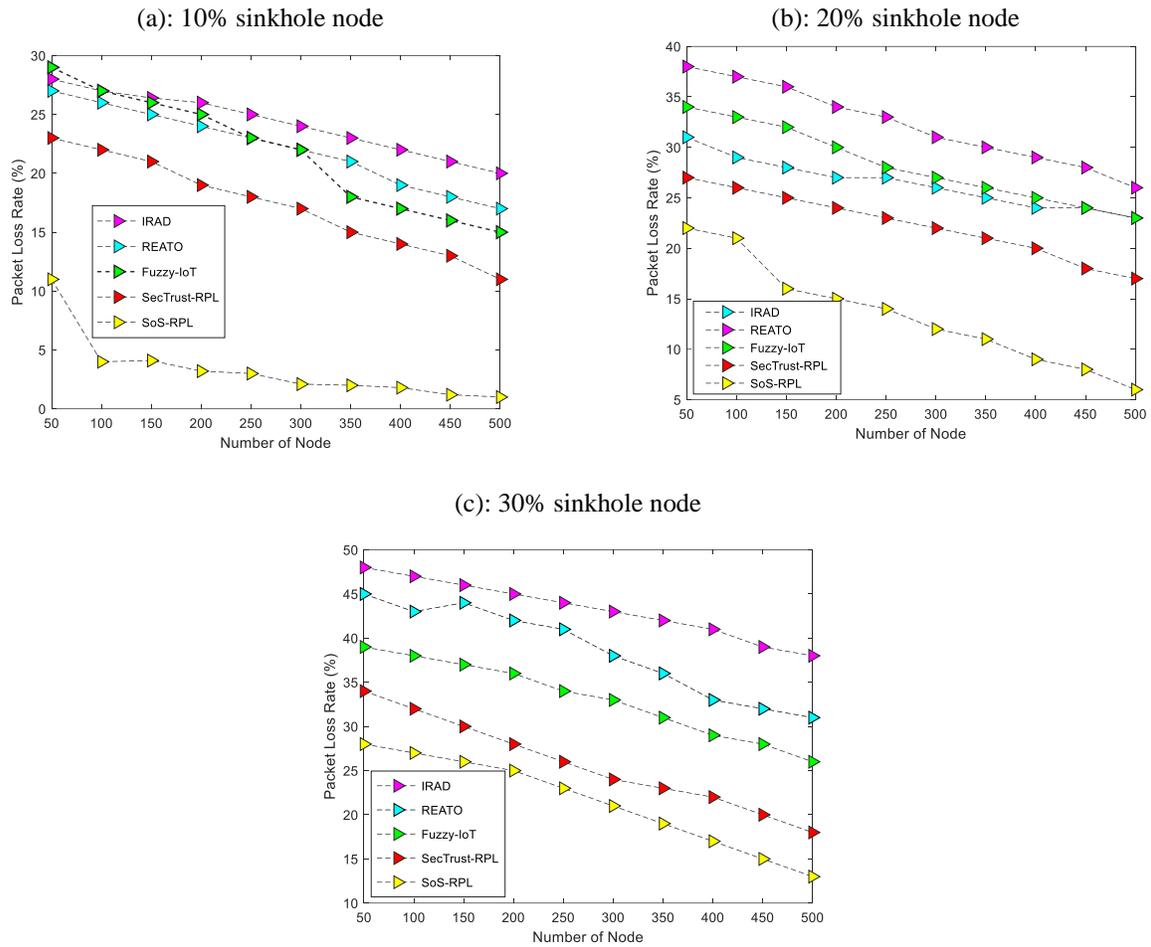

**Fig. 11** PLR vs Number of nodes.

## 5 Conclusion

In this paper, the issue of designing a sinkhole detection method in the internet of things was addressed. In the proposed SoS-RPL method, a sinkhole detection mechanism based on the RPL routing protocol and without the need for any additional hardware was presented. The proposed SoS-RPL method is focused on the IPV6 access of the devices in the IoT network. The malicious nodes can generate fake DIO messages to avoid being detected. However, in the proposed method, these messages are detected and the malicious node is identified and added to the blacklist. Using NS-3, SoS-RPL scheme's performance was analyzed, and it was indicated that it has a high-level performance with a low FPR (below 4.16%), low FNR (below 4.04%), low PLR (below 2.16%), a high level of security, high detection rate (above 96.19%), high PDR (above 94.17%), and high throughput (above 2700 kbps) in comparison with the present methods.

## Conflict of Interest

None.